\documentclass[%
reprint,
superscriptaddress,
onecolumn,
notitlepage,
bibnotes,
 amsmath,amssymb,
 aps,
pra,
]{revtex4-1}

\usepackage{graphicx}
\usepackage{dcolumn}
\usepackage{bm}
\usepackage[]{units}

\begin{document}

\title[arxiv_v2: A nanofocused plasmon-driven sub-10 femtosecond electron point source]
{A nanofocused plasmon-driven sub-10~femtosecond electron point source}

\author{Melanie M{\"u}ller}
\email{m.mueller@fhi-berlin.mpg.de}
\affiliation{Fritz-Haber-Institut der Max-Planck-Gesellschaft, Faradayweg 4-6, 14195 Berlin, Germany}
\author{Vasily Kravtsov}
\affiliation{Department of Physics, Department of Chemistry, and JILA, University of Colorado, Boulder, Colorado 80309, USA}
\author{Alexander Paarmann}
\affiliation{Fritz-Haber-Institut der Max-Planck-Gesellschaft, Faradayweg 4-6, 14195 Berlin, Germany}
\author{Markus B. Raschke}
\email{markus.raschke@colorado.edu}
\affiliation{Department of Physics, Department of Chemistry, and JILA, University of Colorado, Boulder, Colorado 80309, USA}
\author{Ralph Ernstorfer}
\email{ernstorfer@fhi-berlin.mpg.de}
\affiliation{Fritz-Haber-Institut der Max-Planck-Gesellschaft, Faradayweg 4-6, 14195 Berlin, Germany}

\date{11 December 2015}

\begin{abstract}
Progress in ultrafast electron microscopy relies on the development of efficient laser-driven electron sources delivering femtosecond electron pulses to the sample. In particular, recent advances employ photoemission from metal nanotips as coherent point-like femtosecond low-energy electron sources. We report the nonlinear emission of ultrashort electron wave packets from a gold nanotip generated by nonlocal excitation and nanofocusing of surface plasmon polaritons. We verify the nanoscale localization of plasmon-induced electron emission by its electrostatic collimation characteristics. With a plasmon polariton pulse duration below 8~fs at the apex, we identify multiphoton photoemission as the underlying emission process. The quantum efficiency of the plasmon-induced emission exceeds that of photoemission from direct apex illumination. We demonstrate the application for plasmon-triggered point-projection imaging of an individual semiconductor nanowire at 3~$\mu$m tip-sample distance. Based on numerical simulations we estimate an electron pulse duration at the sample below 10~fs for tip-sample distances up to few micrometers. Plasmon-driven nanolocalized electron emission thus enables femtosecond point-projection microscopy with unprecedented temporal and spatial resolution, femtosecond low-energy electron in-line holography, and opens a new route towards femtosecond scanning tunneling microscopy and spectroscopy.
\end{abstract}

\maketitle

\section{\label{sec:Intro}Introduction}
Accessing microscopic phenomena on nanometer length and ultrashort time scales requires probes with corresponding spatio-temporal confinement. In recent years, a variety of microscopy and nanoscale spectroscopy techniques employing electromagnetic radiation from the infrared to the X-ray spectral range as well as electrons have been developed. The spatial and temporal resolution of these techniques is governed by the ability to spatially and temporally confine the corresponding probe pulses.
Ultimately, the transverse and longitudinal extent of a respective wave packet in free space is limited by its wavelength. Whereas optical ultrafast nano-imaging relies on near-field confinement of local {probes\cite{Gramotnev2013, Kawata2009}}, the short wavelength of electrons facilitates free space ultrafast nano-imaging, with spatio-temporal resolution in principle down to angstrom length and attosecond time scales. 
In particular, due to their large scattering cross section and high sensitivity to weak electric and magnetic fields, low-energy electron pulses in the sub-kV range are especially suitable as ultrafast probes for atomic structure in low-dimensional materials as well as nanoscale field distributions. Recently, proof-of-concept time-resolved studies on femtosecond point-projection microscopy (fsPPM) and low-energy electron diffraction have been {realized\cite{Muller2014, Gulde2014, Bainbridge2015}}, utilizing sharp metal tips as ultrafast single electron sources.

The strong field localization around metal nanostructures motivates their application as nanoscale sources for ultrashort electron pulses triggered by femtosecond laser {pulses\cite{Hommelhoff2006, Ropers2007, Yanagisawa2009, Bormann2010, Kruger2011, Dombi2013, Nagel2013, Hommelhoff2015, Greig2016}}. In particular, photoemission from sharp metallic tips has intensly been investigated in recent {years\cite{Hommelhoff2006, Ropers2007, Yanagisawa2009, Bormann2010, Kruger2011, Hommelhoff2015}}. Due to the spatially confined emission and the resulting large transverse coherence length of the {photoelectrons\cite{Ehberger2015}}, such laser-triggered nanotips proved to be ideal point-like sources of high-brightness coherent femtosecond electron wave {packets\cite{Paarmann2012, Quinonez2013}}. Currently, fsPPM is realized with compact setups, where excitation of the sample and photoemission of probe electrons from a nanotip is achieved with two tightly focused laser {pulses\cite{Muller2014, Quinonez2013, Bainbridge2015}}. This allows for a minimal tip-sample distance in the range of 10-20~$\mu$m given by the spatial separation of the laser pulses, and a resulting spatio-temporal resolution of tens of nanometers and tens of femtoseconds, {respectively\cite{Paarmann2012, Muller2014}}. The achievable time resolution, however, is limited by the dispersive broadening of the single electron wave packets during propagation from tip to {sample\cite{Paarmann2012}}. In addition, increased spatial resolution down to 1\,nm or less can in principle be achieved by recording in-line holograms, requiring, however, tip-sample distances below {1\,$\mu$m\cite{Fink1990, Longchamp2013, Longchamp2015}}. This strongly motivates the generation of femtosecond electron wave packets from the apex without direct far-field diffraction limited laser pulse illumination enabling further minimization of the tip-sample distance.

Adiabatic nanofocusing of surface plasmon polaritons (SPPs) provides the spatial confinement of light far below the diffraction {limit\cite{Babadjanyan2000, Stockman2004, Ropers2007a, Berweger2011, Gramotnev2013, Hommelhoff2015}}, enabling ultrafast nanoscale spectroscopy at optical frequencies. A particularly useful implementation of this concept is based on conical gold tapers, where propagating SPPs are launched by illumination of grating structures\cite{Raether1988} and subsequently get nanofocused at the tip {apex\cite{Ropers2007a, Neacsu2010, Berweger2011, Schmidt2012}}. Like a waveguide, the tip transforms the excitation into a confined mode volume, where 10~nm spatial and 10~fs temporal confinement of the plasmonic near fields have been {demonstrated\cite{Berweger2011, Schmidt2012}}. The maximum group delay dispersion experienced by the nanofocused light is found to be on the order of 25\,fs$^2$ for SPP propagation distances between {20-30\,$\mu$m\cite{Schmidt2012, Kravtsov2013}}, supporting broadband SPP coupling and potentially near single-cycle control of the nanofocused field.
The strong spatio-temporal confinement of the evanescent plasmon field allows for generating peak intensities sufficiently high to drive nonlinear processes such as second-harmonic generation\cite{Berweger2011, Schmidt2012, Shahbazyan2013} or four-wave {mixing\cite{Kravstov2015}}. In particular, it has been suggested\cite{Berweger2012} and recently demonstrated\cite{Vogelsang2015, Schroeder2015} that plasmonic nanofocusing can drive nonlinear electron emission from the apex of a nanotip, building on the earlier demonstration of propagating SPP induced electron emission on flat surfaces termed {'plasmoemission'\cite{MeyerzuHeringdorf2015}}. 

Here, we report the nonlocal generation of ultrashort electron wave packets from the apex of a gold nanotip by adiabatic nanofocusing of ultra-broadband SPPs with sub-8~fs duration at optical wavelength and MHz repetition rates. SPPs are generated by broadband grating coupling of 5\,fs optical laser pulses using a chirped grating {design\cite{Berweger2011}}. We use the distinctive collimation properties of the electron beam to characterize the nanofocused SPP--driven electron emission. Nonlinear nanofocused plasmon-driven emission of single-electron wave packets from the apex occurs even for laser pulse energies below 1\,pJ. We find that multiphoton photoemission is the dominant emission process and that the nonlinear electron emission is triggered within a time window of approximately 5 femtoseconds by the nanofocused near-field. We demonstrate the application for fsPPM by imaging the nanoscale surface electric field of a single doped InP nanowire at a tip-sample distance of 3\,$\mu$m, substantially shorter compared to previous fsPPM studies using direct illumination of the tip {apex\cite{Quinonez2013, Muller2014}}. We estimate an electron pulse duration at the sample below 10\,fs for future fsPPM experiments at such small distances. 
In view of its application for fsPPM, an ultrafast electron point source driven nonlocally by nanofocused SPPs lifts the constraint of restricted tip-sample distances and promises improved spatio-temporal resolution. It will further enable the implementation of in-line low-energy electron holography\cite{Fink1990, Beyer2010, Longchamp2013, Longchamp2015} with femtosecond temporal resolution and provides a new route towards ultrafast scanning tunneling microscopy and {spectroscopy\cite{Gerstner2000, Bartels2004, Terada2010, Dolocan2011, Cocker2013}}.

\section{\label{sec:Results}Results and discussion}

\begin{figure}[tbp]
\begin{center}
\includegraphics[width=0.5\columnwidth]{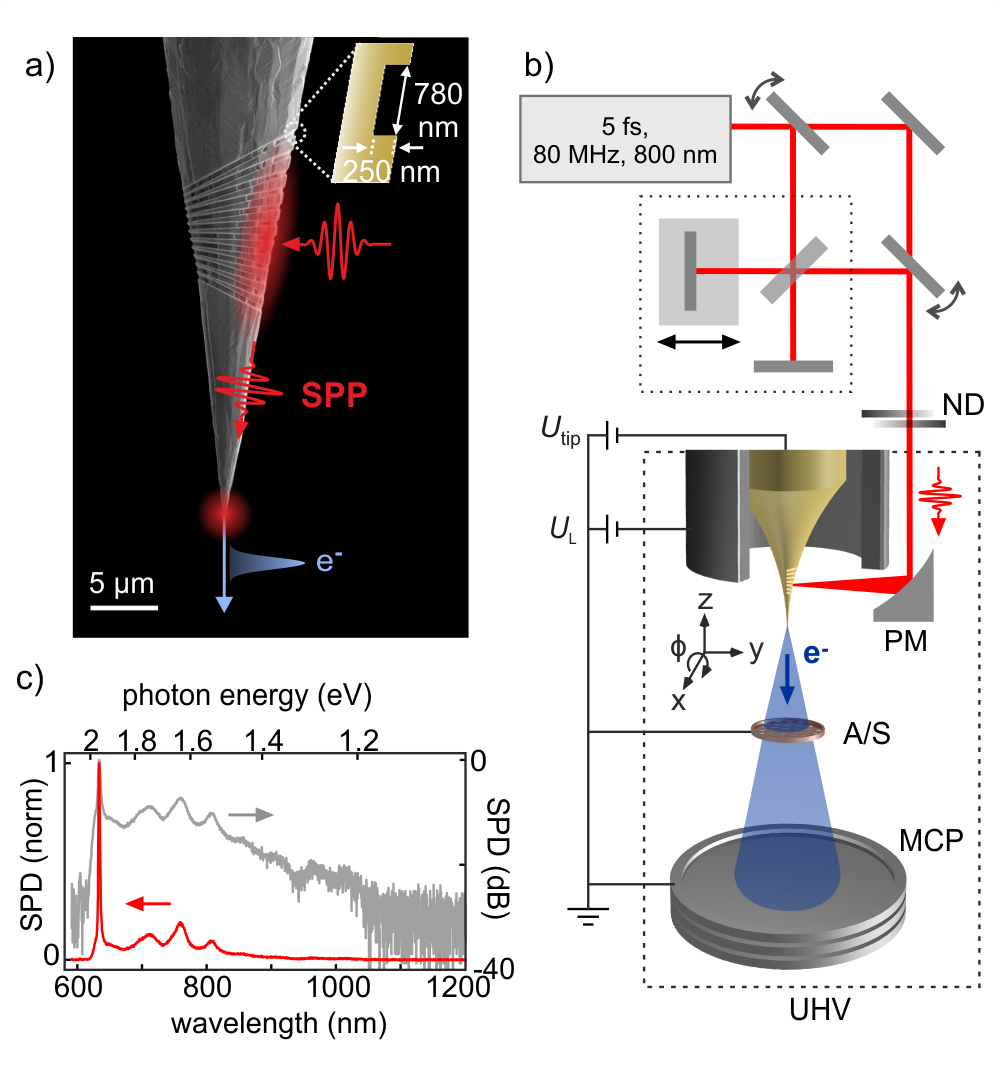}
\caption{Experimental schematic for nanofocused plasmon-induced electron emission from a gold nanotip driven by broadband few-femtosecond laser excitation. a) SEM image of a gold tip with a grating coupler 20 $\mu$m away from the apex with illustration of SPP nanofocusing triggering ultrafast electron emission. b) Corresponding electron pulse imaging setup using an ultrashort 5 fs laser system for plasmon excitation. The biased tips can optionally be mounted inside an electrostatic lens to control the electron beam divergence and the local DC field at the tip. The photoelectrons are accelerated towards the grounded anode hole (or sample) and detected 10 cm behind it. PM: parabolic mirror, ND: neutral density filter, MCP: micro channel plate, UHV: ultrahigh vacuum, A/S: anode/sample. c) Normalized spectral power density (SPD) of the ultra-broadband spectrum of the laser system.}
\label{fig:setup}
\end{center}
\end{figure}

The experimental setup is depicted in Fig.~\ref{fig:setup}b). An ultra-broadband Ti:Sapphire oscillator (Venteon Pulse One) delivers two-cycle pulses at 80\,MHz repetition rate and 2\,nJ pulse energy with its spectrum shown in Fig.~\ref{fig:setup}\,c) and is focused inside an ultrahigh vacuum chamber to a spot with a full width at half maximum (FWHM) of 6\,$\mu$m. Optionally, a Michelson interferometer is used to generate a phase-stable pair of pulse replica with variable time delay.

The tips are etched electrochemically from 125\,$\mu$m polycrystalline gold {wire\cite{Ren2004}}. A grating coupler is cut at 20\,$\mu$m distance from the apex by focused ion beam milling as described {previously\cite{Berweger2011}}. The grating consists of 12 grooves with a center period of 1.5\,$\mu$m linearly chirped for broadband SPP coupling for 90$^\circ$ illumination with tip parallel excitation {polarization\cite{Berweger2011}}. The tips are positioned by a 4-axis manipulator with nm-precision and are optionally mounted inside an electrostatic lens for focusing of the electron {beam\cite{Muller2014}}. The corresponding electron optical system is very similar to that of a Schottky field emission gun in conventional high resolution electron microscopes using virtual point source {cathodes\cite{Orloff2008}}. The tip protrudes the suppressor-type lens by several 100\,$\mu$m which has an outer diameter of 500\,$\mu$m, and negative voltages are applied independently to the tip and the lens. The grounded sample acts as anode and can be positioned with a 6-axis manipulator. Electrons are detected 10\,cm behind the sample with a grounded imaging detector (micro channel plate, fluorescent screen, lens-coupled CMOS camera (Hamamatsu Photonics)). 

We perform 2D and 3D numerical calculations to characterize the focusing conditions of the electron beam, to simulate the projection image of a free-standing nanowire, and to determine the final electron pulse duration at the sample. Details on the simulation procedure can be found in the methods section.\\


\begin{figure*}[htbp]
\begin{center}
\includegraphics[width=1\columnwidth]{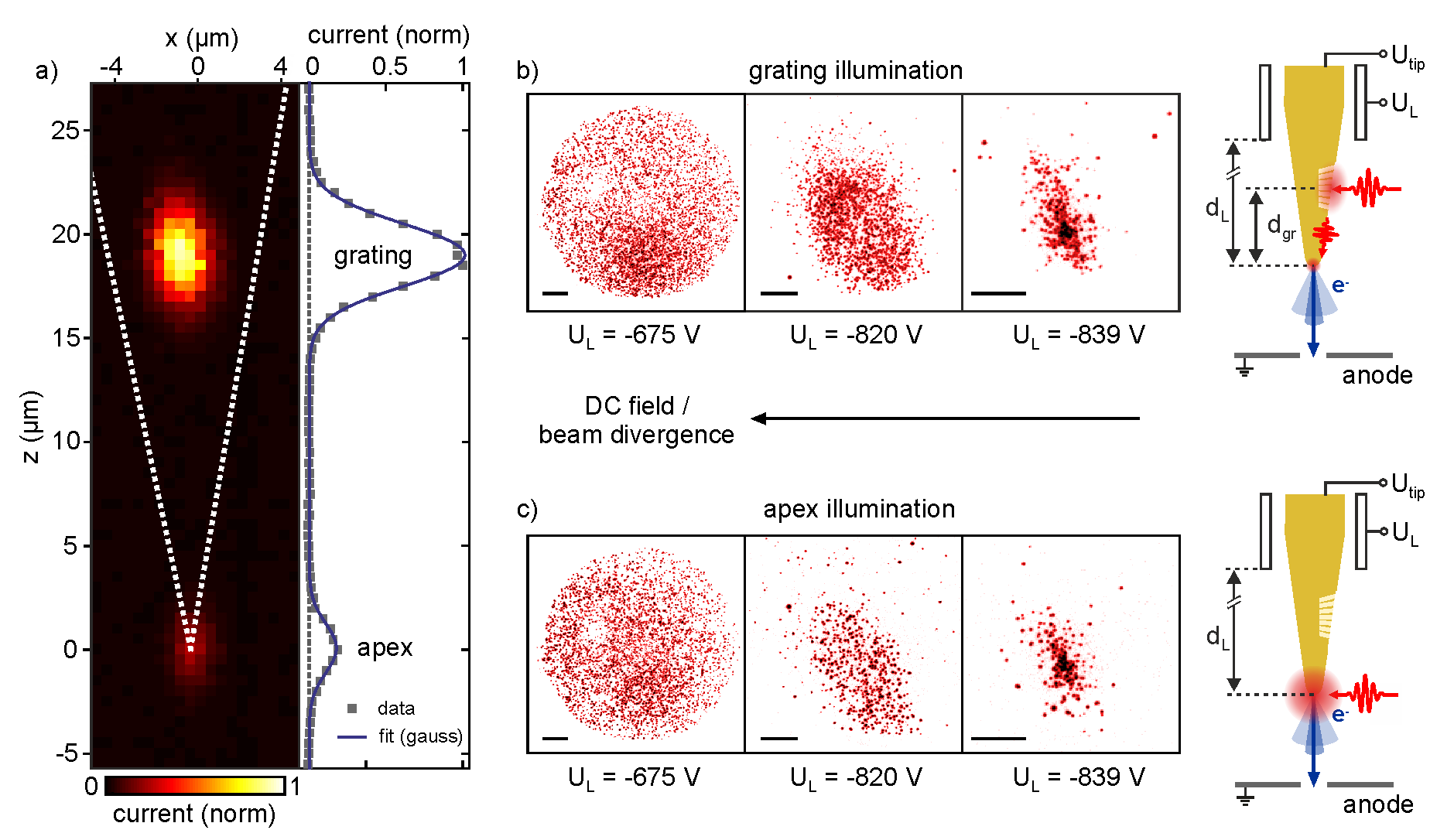}
\caption{Characterization of plasmon-induced electron emission from the apex. a) Spatial map of the photoelectron current recorded at a fluence of $\Phi = 0.5\,\mu$J/cm$^2$ while scanning the tip (white dashed outline) through the laser focus of 6\,$\mu$m width (FWHM) at 1\,s integration for each pixel and at a tip voltage of $U_{\text{tip}} = -150$\,V (no lens was used in this case). The right panel shows the current integrated in the x-direction plotted along the tip axis. Photoelectron emission profiles in the case of grating (b) and apex illumination (c), recorded for a tip placed inside an electrostatic lens for $U_{\text{tip}} = -400$\,V and three different lens voltages $U_{\text{L}} = -675$\,V ($\Phi = 2.4\,\mu$J/cm$^2$, $t_{\text{int}} = 1$\,s), $U_{\text{L}} = -820$\,V ($\Phi = 2.4\,\mu$J/cm$^2$ at grating, $\Phi = 3.2\,\mu$J/cm$^2$ at apex, $t_{\text{int}} = 2$\,s) and $U_{\text{L}} = -839$\,V ($\Phi = 3.7\,\mu$J/cm$^2$, $t_{\text{int}} = 2$\,s) (scale bars are 5\,mm on screen).}
\label{fig:xz-scan}
\end{center}
\end{figure*}

\textbf{Identification of plasmon-driven electron emission.}
The laser-induced electron emission from the nanotip illustrated in Fig.~\ref{fig:setup}a) is first characterized by measuring the photoelectron yield as a function of the nanotip position relative to the laser focus. Figure~\ref{fig:xz-scan}\,a) shows a spatial current map taken for a divergent electron beam emitted from a tip biased at $U_{\text{tip}} = -150$\,V illuminated with laser pulses of 0.6\,pJ energy focused to a fluence of $\Phi = 0.5\,\mu$J/cm$^2$. We observe electron emission when illuminating either the tip apex ($z = 0$) directly or the grating coupler ($z = 19 \,\mu$m).  We find a $\unit{{\sim}4}$x larger electron peak intensity for grating-induced emission compared to direct photoemission from the apex. Notably, within the range of laser intensities employed we observe no photocurrent from other locations along the tip shaft; the right panel in Fig.~\ref{fig:xz-scan}\,a) shows the current integrated along the x-coordinate versus the z-position along the tip's axis.

To verify that the current measured for grating illumination is caused by plasmon-induced emission from the apex, and not by direct photoemission at the grating, we place the tip inside an electrostatic lens and compare the electron beam profiles for direct apex versus grating illumination at different focusing conditions. Figures~\ref{fig:xz-scan}\,b) and c) show the respective emission profiles for three different lens voltages $U_{\text{L}}$ measured at $U_{\text{tip}} = -400$\,V and fluences of $\Phi = 2.4 - 3.7\,\mu$J/cm$^2$. We find very similar spot profiles and collimation voltages for both cases: at a lens voltage of $U_{\text{L}} = -675$\,V the situation is similar to that of a lens-less tip, and photo-excitation generates a divergent electron beam. 
Increasing the lens voltage causes focusing of the electron beams on the detector with comparable spot profiles for both illumination cases, see center and right panels in figures~\ref{fig:xz-scan}\,b) and c) for $U_{\text{L}} = -820$\,V and $U_{\text{L}} = -839$\,V, respectively. As we show in the supporting information experimentally as well as by numerical simulations, there is a direct relationship between the emission site along the tip shaft and the lens voltage required to focus photoelectron wave packets originating from that specific {site\cite{Bormann2015, Schroeder2015}}. Moreover, the spatial distribution of electrons originating from the grating is highly asymmetric due to the one-sided illumination, and is not projected on the detector at low lens voltages (see supporting information for a detailed discussion). Therefore, the agreement in the electron collimation characteristics between illumination of the apex and the grating coupler and the very similar spot profiles are clear evidence for electron emission from the tip apex for both excitation conditions. Additional evidence is provided by the imaging experiments shown below.\\


\begin{figure}[htbp]
\begin{center}
\includegraphics[width=1\columnwidth]{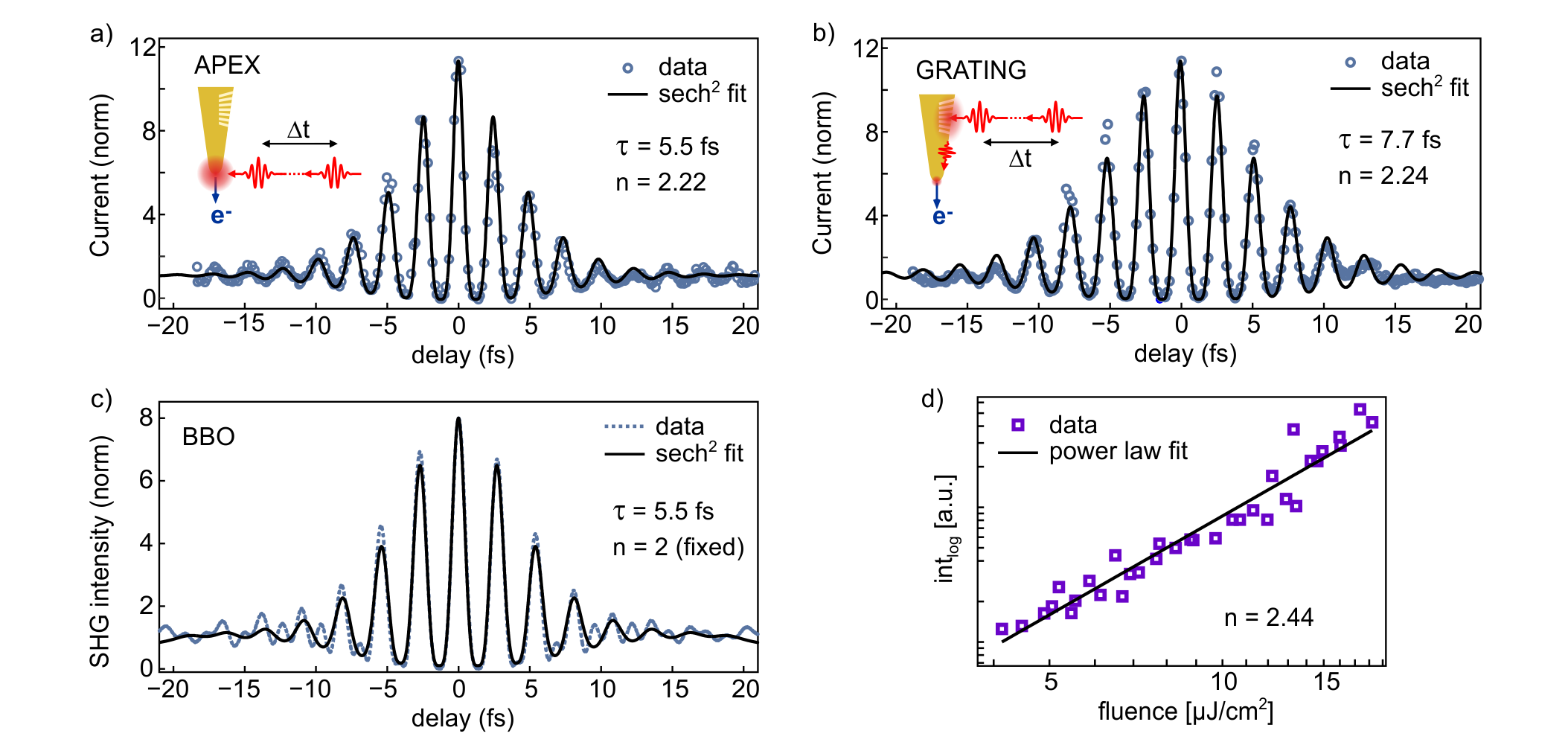}
\caption{Interferometric autocorrelation of the photoelectron current emitted from the apex. IAC measured for apex illumination (a) and by grating coupled SPP-driven photoemission (b). The data (circles) are fitted with a simple squared hyperbolic secant (sech$^2$) pulse shape (black line), revealing pulse durations of $\tau_{\text{ap}} = 5.6$\,fs for direct photoemission and $\tau_{\text{gr}} = 7.7$\,fs for plasmon-driven photoemission, respectively. The order $n$ of the multiphoton photoemission process and the wavelength are also chosen as free parameters in the fit. For comparison, figure (c) shows an interferometric autocorrelation of the incident laser pulse using second harmonic generation (SHG) in a BBO crystal, where a pulse duration of 5.5\,fs is obtained from a sech$^2$ fit. (d) Power dependence of the SPP-driven photocurrent (squares) and corresponding power law fit (solid line).}
\label{fig:IACs}
\end{center}
\end{figure}

\textbf{Temporal characterization of plasmon-driven electron emission.}
We further characterize the temporal profile of the electron emission by two-pulse interferometric autocorrelation (IAC) measurements.
Figures \ref{fig:IACs}\,a) and \ref{fig:IACs}\,b) (open circles) compare the interferometric current from direct photoemission from the apex (a) with that of electron emission from the apex driven by nanofocused SPPs (b), respectively. The IAC from plasmonic nanofocusing is only slightly broadened compared to the IAC obtained from direct apex illumination, indicating that propagating SPPs are generated at the grating coupler with nearly the full laser bandwidth and nanofocused into the apex without significant temporal broadening.

The data is analyzed by fitting autocorrelation functions assuming squared hyperbolic secant (sech$^2$) pulse shapes with pulse duration $\tau$ defined as the FWHM of the intensity profile, center frequency $\nu_0$, order $n$ of the emission process and assuming a flat spectral phase. The electron emission data is fitted as superposition of second- and third-order processes~\cite{Ropers2007, Yanagisawa2011},
\begin{align}
  \text{I}_{ac}(\Delta t) \propto \int^{\infty}_{-\infty} c_2 \left|(E(t) + E(t-\Delta t))^2\right| ^2 + c_3 \left|(E(t) + E(t-\Delta t))^3\right|^2 dt,
\end{align}
with the oscillating electric field $E(t) = \text{sech}(\frac{1.76\cdot t}{\tau})\cdot e^{i 2\pi\nu_0 t}$ and with the order $n = c_2\cdot2 + c_3\cdot3$ being the weighted sum of both contributions where $c_2+c_3 = 1$. 
In the case of direct apex illumination, we obtain a fitted pulse duration of $\tau_{\text{ap}} = 5.5 \pm 0.2$\,fs, in agreement with an IAC measurement of second harmonic generation (SHG) in a BBO crystal at the same position, see Fig.~\ref{fig:IACs}\,c). This agreement implies the absence of localized plasmon resonances at the apex overlapping with the laser spectrum
as this would manifest itself in temporal broadening of the IAC {signal\cite{Lamprecht1999,Anderson2010}}. 
For grating illumination, we retrieve a duration of $\tau_{\text{gr}} = 7.7 \pm 0.3$\,fs of the nanofocused near-field driving the electron emission at the tip apex, which corresponds to three optical cycles and is limited by the coupling bandwidth or propagation dispersion of the SPPs.

We retrieve very similar orders of $n_{\text{ap}} = 2.22 \pm 0.02$ and $n_{\text{gr}} = 2.24 \pm 0.02$ of the electron emission for apex and grating illumination from the IAC fits.
Considering the work function of gold of $\unit{{\sim}5}$\,eV and photon energies centered at $1.6-1.7$\,eV, one would expect a higher order nonlinearity closer to $n = 3$. However, the data shown here is measured for the same tip used in Fig.~\ref{fig:xz-scan}\,b) and~\ref{fig:xz-scan}\,c) at $U_{\text{L}} = -675$\,V with a strongly divergent electron beam, i.e.,~comparably large DC fields on the order of GV/m are present at the apex resulting in a large Schottky effect up to 1\,eV reducing the effective work {function\cite{Gomer1961}}. With increasing lens voltage, i.e.,~with lower DC field strength at the apex, we observe that the order $n_{\text{gr}}$ increases up to a value of 2.6, while the retrieved duration of the nanofocused near-field remains unchanged. We confirm the order of the emission process by measuring the dependence of the photocurrent on the laser fluence incident on the grating. Fig.~\ref{fig:IACs}\,d) shows the power scaling of the SPP-driven photocurrent on a double logarithmic scale measured for a divergent electron beam yielding a comparable order $n_{gr} = 2.44$.

The laser fluences applied to the tip correspond to free-space peak intensities of approximately 4$\cdot 10^8$\,W/cm$^2$. Assuming a field enhancement factor of $k=10$, which is a typical value for gold {tips\cite{Ropers2007}}, we estimate the Keldysh parameter $\gamma$ to be $\approx 35$ in the case of direct apex illumination. With $\gamma \gg 1$, the direct photoemission from the apex occurs in the multiphoton emission {regime\cite{Keldysh1964}}, i.e.,~no optical field effects are expected to contribute to the current. 
We observe that the two-pulse photocurrent away from temporal overlap equals the sum of the individual signals from both pulses, indicating that thermionic electron emission is insignificant.
The efficiency of plasmon-induced electron emission exceeds the direct photoemission fourfold, implying that the losses at the incoupling and during propagation of the SPP are slightly over-compensated by the nanofocusing effect. Nonetheless, the strength of the optical near-fields are of the same order for both excitation schemes, implying that the plasmon-induced electron emission occurs in the multiphoton regime. 
For multiphoton electron emission, the temporal width of the emission probability is $\sqrt{n}$-times shorter than the fundamental intensity envelope. We therefore estimate that the SPP-driven electron emission occurs within a time window of 5\,fs.\\

\textbf{Efficiency of the SPP-driven electron emission.}
At low electron count rates, the SPP-induced electron current can be quantified with the electron imaging detector. The current obtained by illumination of the grating with 0.6~pJ laser pulses, see Fig.~\ref{fig:xz-scan}\,a), is on the order of 2\,fA emitted into a solid angle of 0.032\,sr, which is the field of view of the electron detector. With these low excitation conditions, on average $1.5\cdot 10^{-4}$ electrons are emitted per laser pulse, which corresponds to a quantum efficiency of approximately $5\cdot 10^{-11}$ for the conversion of photons impinging the grating to electrons emitted from the apex.
Taking into account the nonlinearity of the emission process, we extrapolate that 1 electron$/$pulse is emitted when 30~pJ laser pulses are employed, i.e.~with moderate average powers on the order of 2-3\,mW at 80\,MHz repetition rate. In this excitation regime, we estimate the overall quantum efficiency to approach $10^{-8}$.\\


\begin{figure}[htbp]
\begin{center}
\includegraphics[width=1\columnwidth]{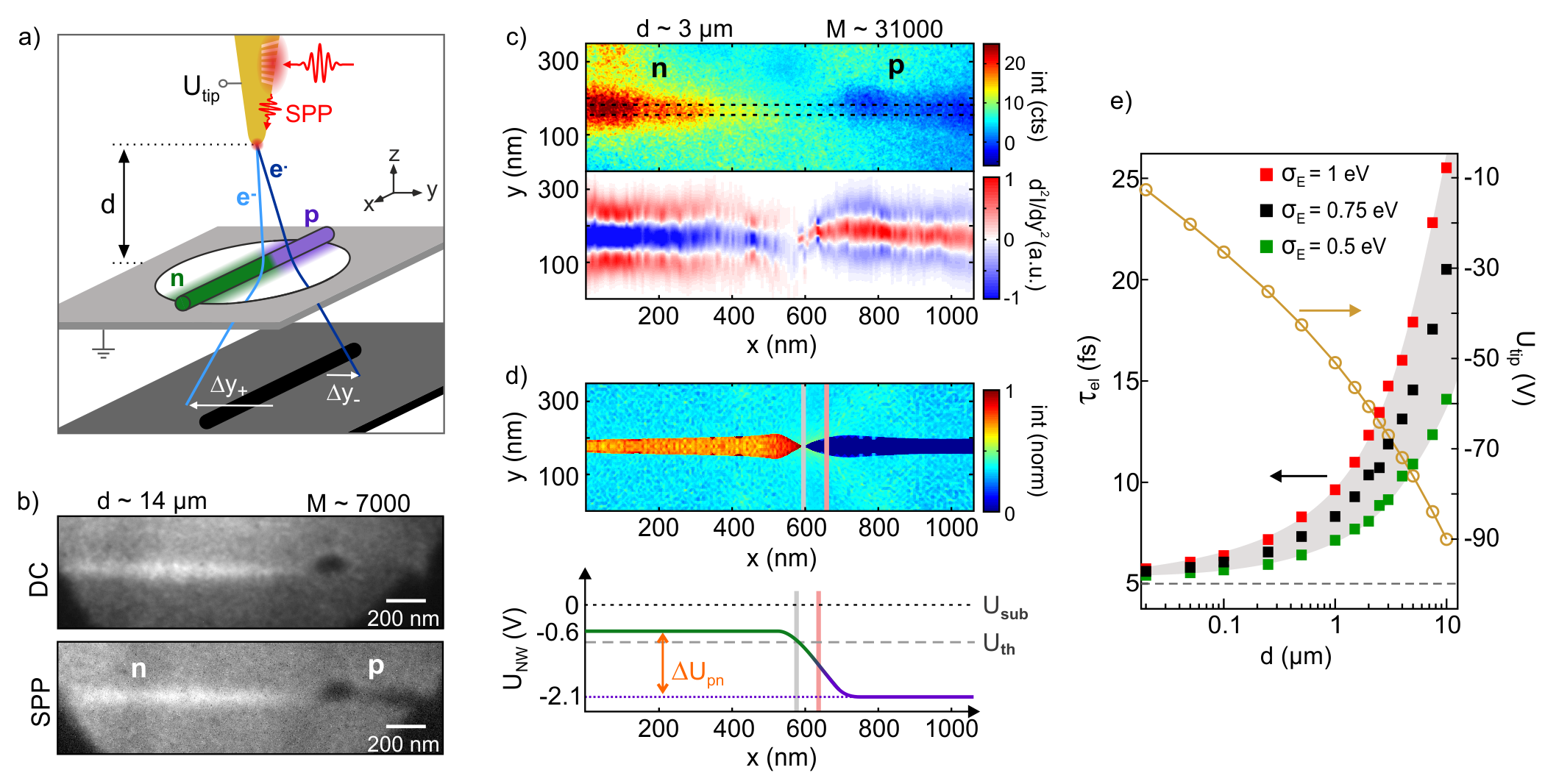}
\caption{Plasmon-driven fsPPM of an individual InP nanowire. (a) PPM schematic for imaging a nanowire with axially varying doping segments. Electron trajectories are deflected by $\Delta y_{\pm}$ depending on the local electric fields. (b) PPM images in field emission mode (top, $U_{\text{tip}} = -126$\,V) and nanofocused SPP-driven mode (bottom, $U_{\text{tip}} = -108$\,V, with full suppression of DC field emission, $\Phi = 3.9\,\mu$J/cm$^2$) at $d = 14\,\mu$m. The change from bright to dark projection reveals the doping contrast. (c) Background-subtracted fsPPM image of the transition region at $d = 3\,\mu$m, corresponding to a geometric magnification of $M \approx 31,000$, in nanofocused SPP-driven mode (top, $U_{\text{tip}} = -60$\,V, $\Phi = 5.5\,\mu$J/cm$^2$). The second derivative of the intensity profile along the $y$-direction (bottom) emphasizes the doping contrast. (d) Simulated PPM image of a 30\,nm wire at $d = 3\,\mu$m (top). The doping contrast is modeled by a potential distribution $U_{\text{NW}}(x)$ (bottom). $U_{\text{th}}$ indicates the threshold from dark to bright projection located $\unit{{\sim}65}$\,nm away from the step center. (e) Simulated FWHM electron pulse duration $\tau_{\text{el}}$ at the sample in dependence of the tip-sample distance. Three initial energy distributions $\sigma_E$ of the electrons are considered, see legend. At each distance, the tip voltage (yellow circles) is scaled to maintain a constant electric field $E_z$ of 1\,GV/m at the apex.}
\label{fig:PPM}
\end{center}
\end{figure}

\textbf{Plasmon-triggered femtosecond point projection microscopy.}
We employ the nanofocused plasmon-triggered electron source for imaging of an individual InP nanowire (NW) by fsPPM to demonstrate its suitability for time-resolved microscopy applications. The NW consists of a $p$- and $n$-doped {segment\cite{Borgstrom2008}}, has a 30\,nm diameter and is spanned across a 2\,$\mu$m hole in a carbon substrate. As illustrated in Fig.~\ref{fig:PPM}\,a) and explained in detail {previously\cite{Muller2014}}, the trajectories of the electrons are strongly influenced by local fields in the vicinity of the NW. Depending on the sign of the electric near-fields, the electrons are deflected in the $x$-$y$~plane with the dominant deflection $\Delta y_{\pm}$ occurring normal to the wire axis towards or away from the NW. In general, the sample-induced displacement of an electron at the detector is directly proportional to the cumulative electric field experienced along its trajectory in the near-field of the {sample\cite{Muller2014}}, i.e.,~$\Delta x \propto E_{x,\text{sample}}$ and $\Delta y \propto E_{y,\text{sample}}$. The point-projection microscopy image is therefore primarily a measure of the electrostatic near-field rather than a shadow image of the geometric structure of the nanoobject. In particular, the point-projection image is sensitive to the doping profile in {nanowires\cite{Hjort2011,Muller2014}}.

Figure~\ref{fig:PPM}\,b) compares PPM images of a NW recorded in DC field emission mode without laser (top) and SPP-driven mode (bottom) with the same tip-sample distance $d = 14\,\mu$m, corresponding to a geometric magnification of $M \approx 7000$. The similarity of the projection images for both electron emission modes provides additional evidence that the electrons emitted by illumination of the grating originate from the apex. The change in the image contrast from bright to dark along the NW axis, i.e.,~from a focusing to a defocusing effect of the NW on the electrons, reveals the change in doping on either sides of the $p$-$n$ junction.

In contrast to previous fsPPM experiments\cite{Muller2014} which were limited to a tip-sample distance of approximately 20~$\mu$m, the nanofocused SPP-driven electron source allows for reducing this spacing without affecting the sample by the electron excitation laser pulse. Fig.~\ref{fig:PPM}\,c) (top) shows a background-subtracted image of the $p$-$n$ junction recorded with the non-locally driven tip at $d = 3\,\mu$m, corresponding to $M \approx 31,000$. For both segments we find constant projected diameters far away from the $p$-$n$ junction, reflecting homogeneous field distributions in these regions. In contrast, the sign reversal of the projected diameter between the segments indicates a strongly inhomogeneous field close to the $p$-$n$ junction. As a comparison, the black dashed lines indicate the real-space diameter as it would be projected in the absence of any electrostatic fields at the sample. The sensitivity of PPM to the doping-dependent near-field along the NW surface is even more apparent in the second derivative $\partial^2 I(x,y)/\partial y^2$ of gaussian intensity profiles $I_{x}(y)$ fitted along the spatial coordinate normal to the wire axis, which is plotted in the lower panel in Fig.~\ref{fig:PPM}\,c). The spatial resolution in the images shown here is limited to a few 10 nanometers by mechanical vibrations. We emphasize, however, that spatial contrast in PPM images of electrostatic fields is not an instrumental-only quantity, as the contrast in the PPM images is determined by the combined geometric and electrostatic properties of tip, sample and substrate.\\
To demonstrate the sensitivity of fsPPM to nanoscale field distributions, we numerically simulate the PPM image of a 30\,nm NW at $d = 3\,\mu$m and $U_{\text{tip}} =-60$\,V (see methods for details). The work function variation along the NW surface caused by the doping profile is modeled by a potential distribution $U_{\text{NW}}(x)$ along the wire axis. Fig.~\ref{fig:PPM}\,d) (top) shows a simulated PPM image for a constant offset bias $U_{\text{NW,off}} = -2.1$\,V with respect to the grounded substrate and a potential step $\Delta U_{pn} = 1.5$\,V centered at $x = 670$\,nm, as illustrated in the lower panel. Whereas these parameters are not adjusted to obtain quantitative agreement between experiment and simulation, the qualitative agreement illustrates the sensitivity of PPM to electrostatic fields on the nanoscale through the electrostatic biprism {effect\cite{Mollenstedt1956}}.

\textbf{Temporal resolution.}
We now estimate the temporal resolution achievable in fsPPM employing SPP-triggered nanotips. The propagation of single electron wave packets and their dispersion between nanotip and a sample plane is simulated for tip-sample distances $d$ between 20~nm and 10\,$\mu$m. 
The numerical methods have been described previously\cite{Paarmann2012} and are briefly summarized in the methods section. In Fig.~\ref{fig:PPM}\,e), the on-axis electron pulse duration $\tau_{\text{el}}$ at the sample as function of $d$ is plotted for three different electron energy distributions with respective energy spreads $\sigma_E$. 
Here, $\tau_{\text{el}}$ is defined as the FWHM of the electron arrival time distribution~\cite{Paarmann2012} convoluted with the temporal profile of the electron emission probability retrieved from the IAC measurement. 
For tip-sample distances in the nanometer range, $\tau_{\text{el}}$ is governed by the electron emission time (gray dashed line), whereas wave packet dispersion is predominant for $d > 10$~$\mu$m. Depending on the energy spread of the electrons, sub-10\,fs electron pulse duration is maintained up to 1-3\,$\mu$m distances from the nanotip.

\begin{figure}[htbp]
\begin{center}
\includegraphics[width=0.5\columnwidth]{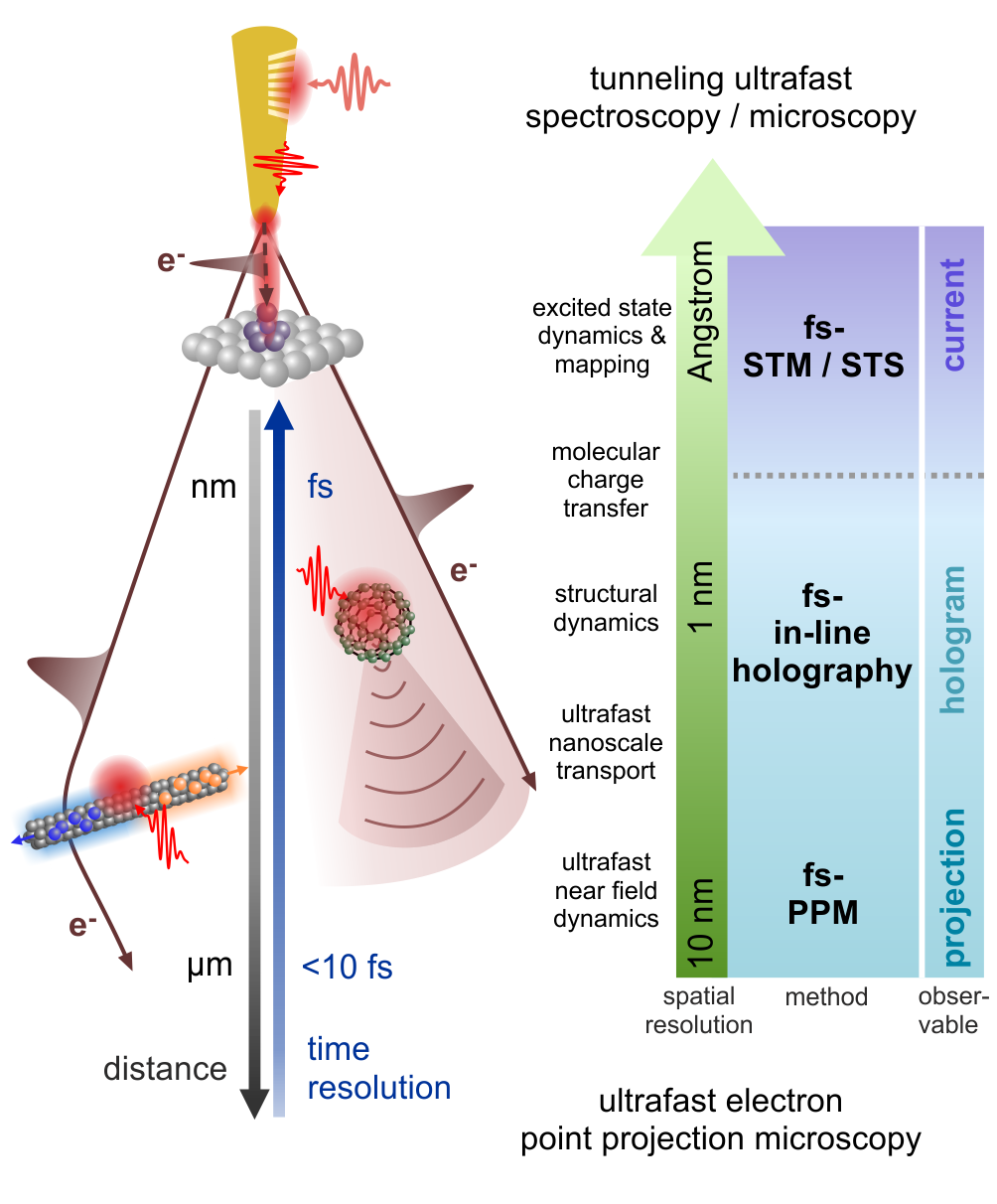}
\caption{Perspectives of nanofocused plasmon-driven ultrafast electron point sources for time-resolved microscopy. The non-local generation of femtosecond low-energy electron pulses enables femtosecond point projection microscopy with a high sensitivity to electromagnetic fields near free-standing nanoobjects, which resembles a non-contact local probe of photocurrents\cite{Muller2014}. With decreasing tip-sample distance in the sub-micrometer range, the projection images transform into holograms, allowing for time-resolved low-energy in-line holography of single molecules, potentially with few-femtosecond temporal and 1 nanometer spatial resolution. With the tip-sample junction entering the the sub-nm range, few-cycle nanofocused SPPs may potentially be employed in time-resolved scanning tunneling microscopy and spectroscopy, with tip--sample coupling possibly in the quantum regime.}
\label{fig:perspective}
\end{center}
\end{figure}

\textbf{Perspectives of femtosecond plasmon-driven electron point sources.}
The nonlocal excitation and subsequent nanofocusing of broadband SPPs triggering nanoconfined ultrafast electron emission from the apex is a major step towards increased spatiotemporal resolution in time-resolved point projection microscopy
at unprecedented geometrical magnification. 
As demonstrated {recently\cite{Muller2014}}, fsPPM provides a sensitive probe for ultrafast photocurrents in nanoobjects through time-resolved investigation of the electrostatic biprism effect.
The new plasmonic nanofocused electron source directly extends the time resolution in fsPPM into the sub-10~fs regime.
By reducing the tip-sample distance to the sub-$\mu$m range, the purely geometric projection transforms into a hologram and fsPPM merges into femtosecond low-energy electron in-line holography. \\
In-line holographic imaging of individual biological specimen with 1\,nm spatial resolution at the anode has been realized recently\cite{Longchamp2015, Longchamp2015a} by using graphene\cite{Longchamp2013} as sample support, thus reducing the biprism effect which is detrimental if high spatial resolution is {desired\cite{Weierstall1999}}.
Beyond such sample restrictions, the spatial resolution of femtosecond in-line holography will ultimately be determined by the spatial coherence of the electron source, which is given by the effective source size $r_{\text{eff}}$ and the electron energy {spread\cite{Spence1994}}.
While the transverse coherence properties of ultrashort electron wave packets emitted from nanotips have not yet been thoroughly investigated, an effective source size of $<1$\,nm comparable with values for DC field emission was found for linear photoemission from tungsten {tips\cite{Hommelhoff2015}}. Moreover, the energy spread of ultrashort electron wave packets is ultimately given by their Fourier limit, which amounts to approximately 500~meV for 5~fs pulses.
Hence, as the concepts of low-energy electron holography are compatible with ultrafast nanofocused SPP-driven electron sources, the prospective combination of sub-10~fs temporal and 1~nm spatial resolutions would enable the investigation of ultrafast charge transport on electronic time and molecular length scales.\\
At tip--sample distances in the sub-nanometer range, SPP-driven electron point sources are promising probes for time-resolved scanning tunneling microscopy (STM), which has been pursued for more than two decades\cite{Hamers1990,Gerstner2000,Grafstrom2002,Terada2010} but remains challenging due to laser-induced thermal expansion\cite{Grafstrom1991,Lyubinetsky1997,Gerstner2000a} and contribution of the transient hot electrons to the tunneling {currents\cite{Dolocan2011}}.
The \textit{non-local} excitation of the optical near-field in the tip--sample junction via nanofocused SPPs may help to overcome these limitations. With the junction at the atomic scale ($\unit{{\sim}3}$~\AA), plasmonic tunneling\cite{Zuloaga2009,Esteban2012,Savage2012,Tame2013} can provide a conductance channel with high degree of spatial confinement and potentially ultrafast and controllable temporal response for robust femtosecond time-resolved STM. We point out that using SPPs with frequencies in the near-infrared and visible spectral range, as employed here, permits significantly stronger spatiotemporal confinement in the tip--sample gaps both in the classical near-field coupling and quantum tunneling regimes, as compared to IR and THz plasmons.\\
Figure~\ref{fig:perspective} summarizes the potential applications of non-locally driven plasmonic femtosecond electron point sources for ultrafast point projection microscopy, time-resolved electron holography and potentially scanning tunneling microscopy and spectroscopy.

\section{\label{sec:Concl}Conclusion}
We demonstrated photoemission of sub-10 femtosecond electron pulses from the apex of a gold nanotip driven by the nanofocused near field of surface plasmon polaritons generated 20\,$\mu$m away from the apex. Employing the unique ability of adiabatic nanofocusing to confine ultra-broadband few-cycle laser pulses to a nano-sized spot at the apex, we realized a remotely-driven few-femtosecond electron point source operated at high-repetition rates and at optical frequencies. We further performed plasmon-triggered point-projection microscopy of an individual nanowire which allowed for a significant reduction of the tip-sample distance down to $3\,\mu$m. Beside the increase in spatial resolution, future experiments on time-resolved point projection microscopy will greatly benefit from the reduced electron propagation length increasing the temporal resolution to the sub-10\,fs range. Moreover, taking advantage of the large coherence length of low-energy electron wave packets excited from the apex of a nanotip, this will enable the realization of femtosecond in-line holography at sample distances below 1\,$\mu$m. Ultimately, the realization of ultrafast scanning probe techniques employing plasmon-triggered tunneling of femtosecond electron wave packets becomes conceivable.

\section{\label{sec:meth}Methods}
\textbf{Experimental.}
The tip-sample distance in PPM is calculated by moving a defined step $\Delta y$ with the sample and measuring the projected step $\Delta y_{\text{det}} = M \cdot \Delta y$ on the detector. The tip-sample distance is then determined by the magnification and the detector distance as $d = D_{\text{det}} / M$, given that $D_{\text{det}} \gg d$. 

The PPM image at $d = 3\,\mu$m in Fig.~\ref{fig:PPM}\,c) is background corrected by subtracting an image recorded with the same settings but moving the nanowire by 500\,nm out of the detector.  $\partial^2 I(x,y)/\partial y^2$ plotted in Fig.~\ref{fig:PPM}\,d) is obtained by fitting $I(y)$ with Gaussian intensity distributions along the wire axis $x$ after binning of 10 adjacent pixel lines.\\

\textbf{Simulations.}
The numerical simulations shown here and in the supplementary information follow the basic procedure explained in reference~\citenum{Paarmann2012}. We calculate the electrostatic properties of the particular geometry using a finite element method (COMSOL Multiphysics 5.1). Propagation of single electron wave packets in the respective electric field is simulated classically by using a Runge-Kutta algorithm to solve the equation of motion (MATLAB or COMSOL Multiphysics 5.1.).

The PPM image in Fig.~\ref{fig:PPM}\,d) is obtained from a 3D simulation of the electron trajectories from the tip to the detector plane, passing by a 30\,nm nanowire positioned 3\,$\mu$m below the tip. We choose the $x$-$z$ plane spanned by the wire and tip axis at $y = 0$ as symmetry plane to reduce the computational cost. The nanowire doping profile is modeled by a step-like potential distribution along the wire axis as explained in detail in the supporting information of reference~\citenum{Muller2014}. Electrons are emitted normal to the tip surface with an energy of 0.5\,eV and their trajectories are calculated for emission angles between 0$^\circ$ and 19$^\circ$ in steps of 0.14\,$^\circ$. The projection image is then derived from the electrons final position in the detector plane.

To obtain the final electron pulse duration at the sample, we calculate at each tip-sample distance the on-axis electron trajectories for initial gaussian energy distributions centered at $E_0 = 0.1$\,eV with different energy width $\sigma_E$ (see legend in Fig.~\ref{fig:PPM}\,e)). For the propagation we assume prompt electron emission and then convolute the electron arrival time distribution at the sample with the probability distribution for electron emission. The latter is calculated from the intensity profile of a sech$^2$-pulse using the parameters obtained from the grating IAC fit and taking into account the nonlinearity of $n = 2.24$ (giving a FWHM of 5\,fs). We then define the resulting FWHM as the final electron pulse duration.

\section{\label{sec:Ackn}Acknowledgements}
R.E. acknowledges funding from the Max Planck Society. V.K. and M.B.R. acknowledge funding from the National Science Foundation (NSF Grant CHE 1306398), and the work was in part supported by the U.S. Department of Energy, Office of Basic Energy Sciences, Division of Materials Sciences and Engineering, under Award No. DE-SC0008807. We thank Sibel Yalcin for help with tip fabrication and instrument support from the Environmental Molecular Sciences Laboratory (EMSL), a national scientific user facility from the DOE Office of Biological and Environmental Research at Pacific Northwest National Laboratory (PNNL). PNNL is operated by Battelle for the U.S. DOE under the contract DEAC06-76RL01830. We thank M.~Borgstr{\"o}m for providing the nanowire sample.

\section{\label{sec:Suppl}Supplementary Information}
Operation principle of electron optical system, numerical simulations of electrostatic properties and particle trajectories, identification of electron emission sites by comparison of measured and simulated spot profiles from photoemission at apex and grating.


\bibliography{SPP_electron_emission_revision}

\end{document}